\documentclass[graybox]{svmult}

\usepackage{mathptmx}       
\usepackage{helvet}         
\usepackage{courier}        
\usepackage{type1cm}        
\usepackage{makeidx}         
\usepackage{graphicx}        
\usepackage{multicol}        
\usepackage[bottom]{footmisc}

\makeindex             

\begin{document}

\title*{Dwarf galaxies in the nearby Lynx-Cancer void: photometry, colours and ages}
\author{Simon Pustilnik, Alexei Kniazev, Yulia Lyamina and Arina Tepliakova}
\institute{Simon Pustilnik and Arina Tepliakova \at SAO, Nizhnij Arkhyz, Karachai-Circassia, 369167, Russia, \email{sap@sao.ru, arina@sao.ru}
\and Alexei Kniazev \at SAAO, PO Box 9, 7935 Observatory, Cape Town, South Africa, \email{akniazev@saao.ac.za}
\and Yulia Lyamina \at SFU, Rostov-on-Don, Russia, \email{jlyamina@yandex.ru}}
%
%
\maketitle

\vskip-1.2truein

\abstract{
The nearby Lynx-Cancer void is a good laboratory to study the effect
of very rarefied environment on the evolution of the least massive dwarf
galaxies.
A recently compiled sample of this void's galaxies includes about one hundred
objects with M$_{\rm B}$ in the range --12 to --18 mag.
Good quality images are available in the SDSS database for $\sim$80\%
of the sample. Their $u,g,r,i,z$ photometry allows one to derive galaxy
stellar mass
(and, incorporating HI data, gas mass-fraction) and ages of visible stellar
populations, and hence, the epoch of their formation (first SF episode).
We present the first photometric results of the ongoing study of the
Lynx-Cancer void.}

\section{Objectives}
\label{sec:1}
\begin{figure*}
 \centering
\includegraphics[width=4.0cm,angle=-90]{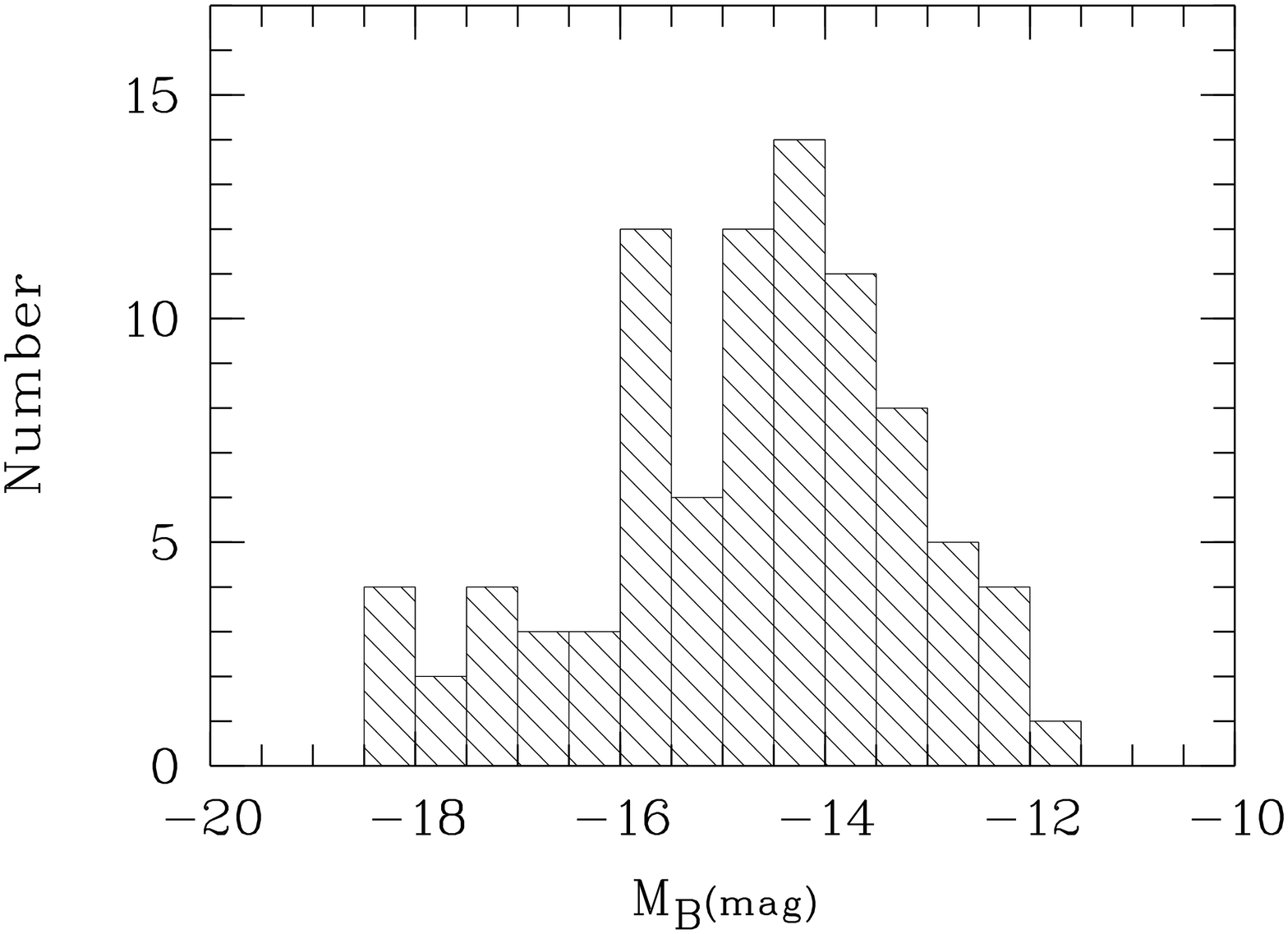}
\includegraphics[width=4.0cm,angle=-90]{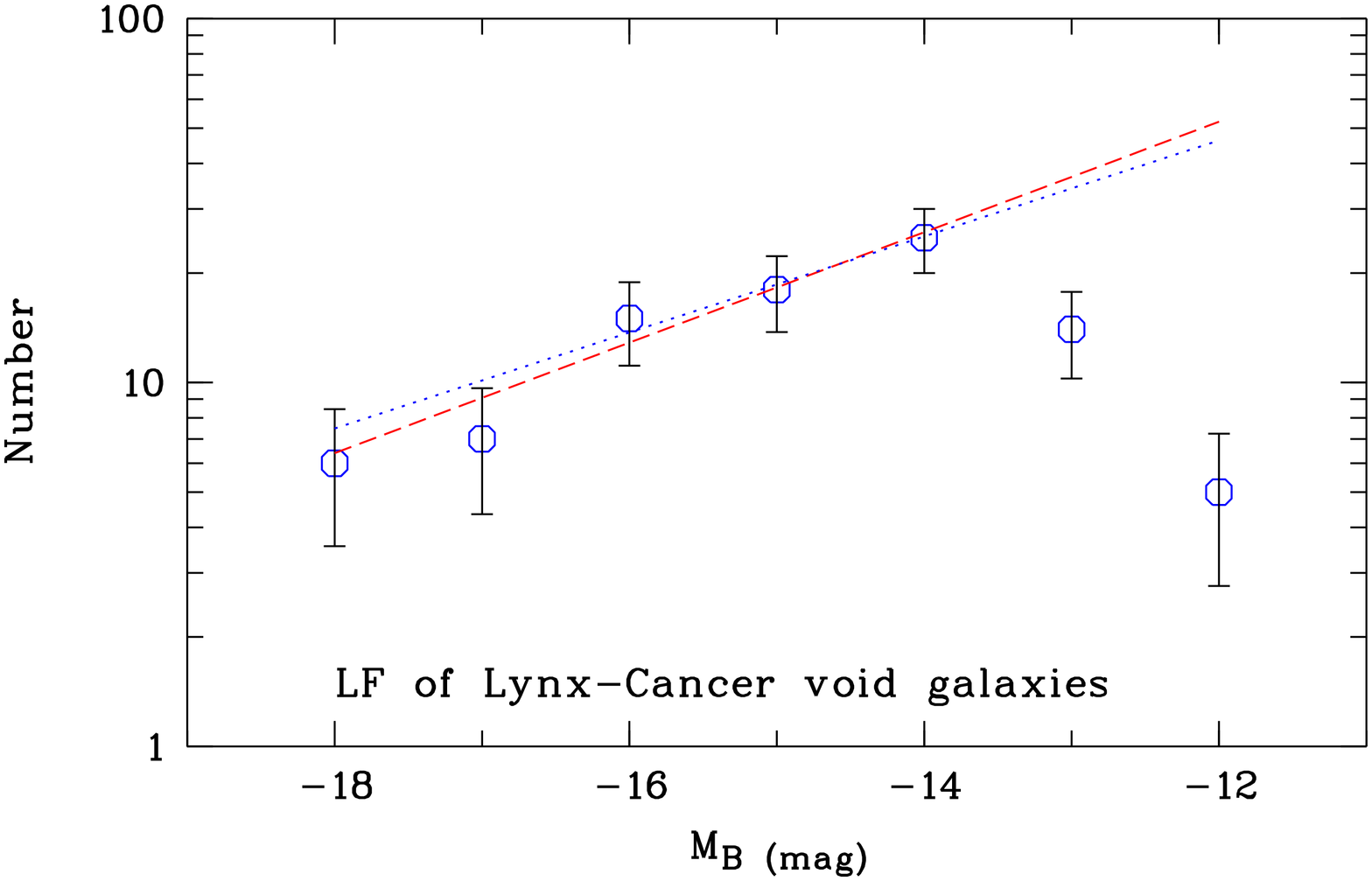}
\caption{\label{fig:Pustilnik1}
{\it Left:} The distribution of absolute blue magnitudes M$_{\rm B}$ for
Lynx-Cancer void galaxies. {\it Right:} The shape of the raw luminosity
function of this void galaxies. The two lines show fits by Schechter function
on the range M$_{\rm B}$ of --14.0 to -18.0 mag for two adopted  values
of M$_{\rm B}^{*}$: the standard: --20.2~mag. (dotted, $\alpha$=--1.33), and
the reduced value, suitable for voids, --19.2 mag. (dashed, $\alpha$=--1.38).
The Hubble constant is adopted 73~km~s$^{-1}$~Mpc$^{-1}$. }
\end{figure*}

The best probes of possible effects of environment on galaxy evolution and
formation should be the most fragile, least massive dwarfs. We compiled the
largest and deepest sample of dwarfs, falling within the nearby Lynx-Cancer
void (\cite{SA0822}, Pustilnik \& Tepliakova, 2010, MNRAS, submitted):
about a hundred objects with M$_{\rm B}$ down to --12 mag. (see left panel
of Fig.~\ref{fig:Pustilnik1} for the void galaxies' M$_{\rm B}$
distribution).

The main goal of this study is to look for the 
evolutionary parameters of the void's galaxies and compare them with
parameters of their counterparts
in denser environment. For this we perform the spectroscopy of HII regions
in void galaxies to derive the oxygen abundances.  We also observe these
galaxies in HI 21-cm line to derive their total gas masses.
Combining this data with SDSS photometry we obtain the evolutionary
parameter f$_{\rm gas}$ -- the ratio of gas mass to the whole baryonic mass.
As a part of the ongoing project, we examine
$u-g,g-r,r-i$ colours of outer parts of the void galaxies from their SDSS
images. Comparing them with PEGASE model evolutionary tracks, we derive
estimates on ages of stellar populations. We also analyse the Surface
Brightness (SB) profiles of void galaxies \cite{K04}, to obtain the
distribution of $\mu_{0,B}^{ext,i}$, $B$-band central SB corrected for
the Galaxy extinction and inclination.

\section{Results}
\label{sec:2}

\subsection{Colours}
\begin{figure}[t]
\includegraphics[angle=-90,width=9.5cm]{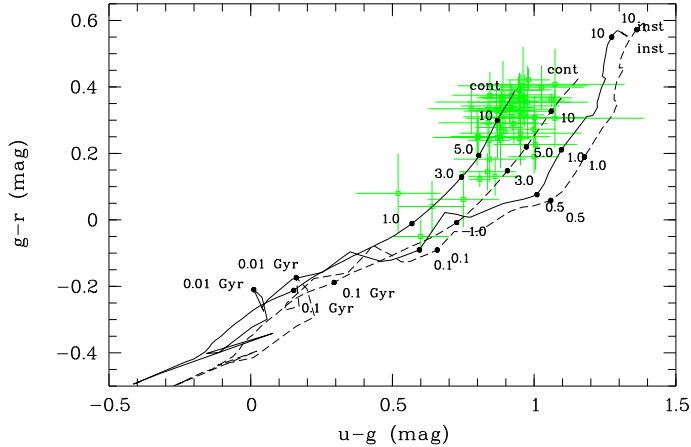}
\caption{%
The $ugr$ colours of the outer parts of 47 Lynx-Cancer void galaxies are
shown in comparison to the PEGASE evolutionary tracks for two IMFs -
Salpeter (solid lines) and Kroupa (dashed lines) and two extreme SF laws
- instantaneous and continuous with constant SFR.
Numbers along the tracks indicate the time elapsed since the beginning of
SF episode. The great majority of the void galaxies show $ugri$ colours
consistent
with the cosmological ages T$\sim$10--15~Gyr typical of other galaxies.
Only four galaxies show no traces of stars with T$>$1--3~Gyr. Five more
galaxies show colours for stars with intermediate ages, of
T$\sim$4--7~Gyr.
}
\label{fig:Pustilnik2}
\end{figure}

In Fig.~\ref{fig:Pustilnik2},  for 47 L-C void galaxies with SDSS images we
show $ugr$ colours of their outer parts in comparison to PEGASE \cite{pegase}
evolutionary tracks for two extreme Star Formation (SF) laws: instantaneous
SF and continuous SF with constant SFR. Both tracks for the standard Salpeter
and the Kroupa \cite{K93} IMFs are used for comparison. The adopted
metallicity parameter, z=0.002, matches the range 0.0004--0.004 typical of
void dwarfs. The great majority of the studied galaxies show a sizable
amount of old stellar populations, corresponding to continuous SF with
ages of T=10--15 Gyr. Only for 4 dwarfs do the outer region colours correspond
to ages T$\sim$1--3 Gyr.
For 5 more galaxies, the oldest visible population has T$\sim$4--7 Gyrs.
Galaxies with "small" ages also appear to be the most metal-poor and gas-rich.

\subsection{LSBG population}
\begin{figure}[t]
\sidecaption
\includegraphics[width=7.5cm]{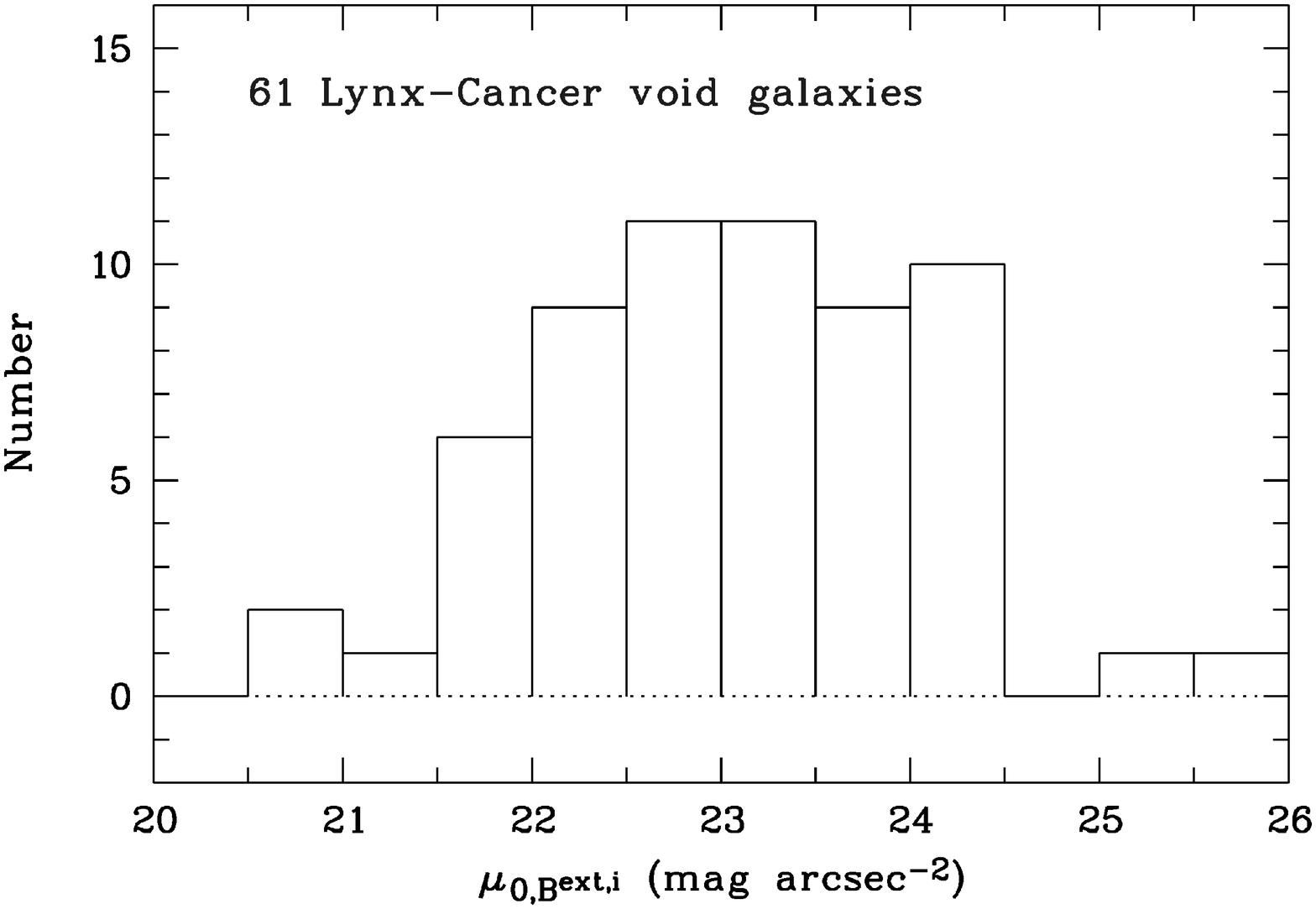}
\caption{%
The distribution of the central SB ($\mu_{0,B}^{ext,i}$) for
61 L-C void galaxies. For galaxies with the sizable central SF
region or ``buldge'' we accepted the central SB of the underlying disc.
About a half of galaxies belong to the low SB regime: $\mu_{0,B}^{ext,i} >$
23 mag~arcsec$^{-2}$. However, many of them passed through the SB threshold
for SDSS spectroscopy due to the significant inclination-related brightening.
Many others have no SDSS spectra, but have the known velosities/distances
either from HI observations or from the photometry of resolved stars.   }
\label{fig:Pustilnik3}
\end{figure}

More than a half of the Lynx-Cancer void galaxies (mainly the faint
end) were identified as a result SDSS spectroscopy.
The latter was limited by the higher SB galaxies. Namely, it missed more
than 50\% of objects with the observed $\mu_{0,B} > 23.2$ mag~sq.arcsec$^{-2}$
(Pustilnik \& Tepliakova, 2010, MNRAS, submitted; as recalculated
for purely exponential discs from the estimates in \cite{Blanton05}). On the
other hand, since $\mu_{0,B}$ is related to galaxy luminosity (e.g.,
\cite{CD02}), one expects that LSBGs could be the main population of voids.
In Fig.~\ref{fig:Pustilnik3}, we show the distribution of $\mu_{0,B}^{ext,i}$
(corrected for the Galaxy extinction and inclination)
for 61 Lynx-Cancer void galaxies examined so far.
Their values of $\mu_{0,B}$ were transformed from the respective $g$
and $r$ values using \cite{L05}.
The fraction of LSBGs (that is, with  $\mu_{0,B}^{ext,i} > 23.0$) in this
subsample appears $\sim 0.5$ and is quite representative of the whole sample
selected this way. Due to the evident selection effects, the significant part
of the void LSB dwarfs remains undisclosed. This is the challenge for
both observers and for CDM N-body simulations of void galaxy population.

\subsection{Luminosity function}

The luminosity function (LF) of void galaxies is an important parameter
for comparing predictions of cosmological models of galaxy and
structure formation. To date, the Lynx-Cancer void galaxy sample
is significantly larger than for any other individual void. Therefore,
even the preliminary information on this raw LF (see
Fig.~\ref{fig:Pustilnik1},Right) is of interest. For M$_{\rm B} >-13.5$,
the significant loss (presumably due to selection effects)
of faint galaxies is seen. However, for M$_{\rm B}$ range of --14.0 to
--18.0, the fitting of LF by Schechter function works well and results in the
power-law index $\alpha \sim$--1.33 to --1.38$\pm$0.06, depending on the
adopted value of M$_{\rm B}^{*}$. This slope is close to that for
the LF of SDSS galaxies from the study \cite{Blanton05}, uncorrected for
the SB selection effect.

\section{Unusual dwarfs in Lynx-Cancer void}
\label{sec:3}
In the course of the study of this void galaxy sample, several of the most
metal-poor galaxies with 12+log(O/H) $\le$ 7.30 were uncovered.
They include SDSS J0926+3343 (7.12) \cite{J0926}, DDO 68 (7.14)
\cite{DDO68,IT07}, J0737+4724 (7.24), J0812+4836 (7.28) \cite{IT07},
J0852+1350 (7.28).
The three former galaxies show no tracers
of stellar populations with T $>$1--3 Gyr. Additionally,
we found two LSBDs (J0723+3622 and SAO 0822+3545) with unknown metallicities
with overall blue colours, indicating ages T $\sim$1--3~Gyr.
The statistical significance of this finding will be addressed in a
forthcoming paper. But such unusual concentration of the most metal-poor
and `unevolved' dwarfs
suggests on a sizable effect of void environment on galaxy evolution.

%
%


\end{document}